\documentclass[10pt]{iopart}
\usepackage{graphicx}
\newcommand{\ZnMnInSe}{{$\mathrm{Zn}_{1-x}\mathrm{Mn}_{x}\mathrm{In}_2\mathrm{Se}_4$}}
\newcommand{\ZnMnInTe}{{$\mathrm{Zn}_{1-x}\mathrm{Mn}_{x}\mathrm{In}_2\mathrm{Te}_4$}}
\newcommand{\bZnMnInSe}{{$\mathbf{Zn_{1-x}Mn_{x}In_2Se_4}$}}
\newcommand{\MnInSe}{{$\mathrm{Mn}\mathrm{In}_2\mathrm{Se}_4$}}
\newcommand{\ZnInSe}{{$\mathrm{Zn}\mathrm{In}_2\mathrm{Se}_4$}}
\newcommand{\Mn}{$\mathrm{Mn}$}
\newcommand{\Zn}{$\mathrm{Zn}$}
\newcommand{\Cu}{$\mathrm{Cu}$}
\newcommand{\Ni}{$\mathrm{Ni}$}
\newcommand{\Se}{$\mathrm{Se}$}
\newcommand{\In}{$\mathrm{In}$}
\begin{document}
\bibliographystyle{unsrtJPCM}
\title[Structure of \ZnMnInSe]{Structure of \bZnMnInSe\ crystals grown by CVT}

\author{J.~Mantilla\dag, G.~E.~S.~Brito\dag, E.~ter~Haar\dag,
V.~Sagredo\ddag, V.~Bindilatti\dag%
\footnote[3]{E-mail: \texttt{vbindilatti@if.usp.br}}}

\address{\dag\ Instituto de F\'{\i}sica, Universidade de S\~{a}o Paulo\\
Cx. Postal 66.318, 05315--970 S. Paulo, SP, Brazil}

\address{\ddag\ Laboratorio de Magnetismo, Universidad de los Andes,
M\'{e}rida 5101, Venezuela}

\begin{abstract}
Single crystals of \ZnMnInSe\ were grown by the chemical vapour
phase transport (CVT) technique. Through X-rays powder diffraction
patterns and Laue diagrams of single crystals we studied the
transformation from the layered rhombohedral structure of \MnInSe\
to the tetragonal structure of \ZnInSe. On the \ZnInSe\ side, we
observe single-phase, solid solution samples for $x{=}0.01$ and
$x{=}0.25$, as is the case for the \MnInSe\ side with $x{=}1$ and
$x{=0.87}$. For the intermediate concentrations $x{=}0.35$,
$x{=}0.60$ and $x{=}0.67$ we observe our samples to be two-phase
mixtures.
\end{abstract}


\submitto{\JPCM}


\section{Introduction}
While much is known about II--VI diluted magnetic semiconductors
(DMS) containing manganese\cite{Dietl94}, the structurally and
magnetically more complex ternary and quaternary systems have
recently begun to be studied\cite{Nikiforov99}, in the expectation
of exploring and manipulating the interactions between the
electronic, magnetic and structural degrees of freedom. An example
is cation disorder, which can be studied by its effect on the
magnetic properties\cite{Wooley95, Moron01}. Spin-glass behaviour
was observed in \ZnMnInTe{\cite{Goya01} and in
\MnInSe\cite{Mantilla04}, end-point of the \ZnMnInSe\ series we
investigate here. The low \Mn\ concentration side of the series is
structurally similar to the II-VI DMS, in which extensive
investigation of the exchange interactions between
$\mathrm{Mn^{2+}}$ ions has been performed\cite{Shapira02}. To
extend these investigations to \ZnMnInSe, an important first
question is for what substitution concentrations it is possible to
obtain homogeneous solid solutions, since in this series the two
endpoint compounds are in different crystal systems.

The structures of the ternary compounds of the
$\mathrm{II{-}III_2{-}VI_4}$ family (II: bivalent metal, III:
trivalent metal and VI: chalcogen atom) are found in three major
types: a cubic structure (spinel), a tetragonal defective zinc
blende structure and a rhombohedral structure \cite{Fiorani83}.
The latter two structures are realized by the end-points of the
\ZnMnInSe\ series.

\ZnInSe\ crystallizes in a tetragonal cell of space group
$I\bar{4}2m$ with parameters $a{=}5.710$~$\mathrm{\AA}$ and
$c{=}11.420$~$\mathrm{\AA}$. This structure is a defective
chalcopyrite with the metal atoms randomly distributed within the
cationic sub-lattice\cite{Gastaldi87,Marsh88}. \MnInSe\ exhibits a
rhombohedral structure with space group $R\bar{3}m$ and lattice
constants $a{=}4.051$~$\mathrm{\AA}$ and $c{=}39.460$~$\mathrm{\AA}$. In
this layered structure, the unit cell consists of three van der
Waals coupled slabs, each slab consisting of four \Se\ layers in
the sequence $ABCA$. Between these layers there are octahedral and
tetrahedral sites, which are again thought to be randomly filled
by the \Mn\ and \In\ atoms\cite{Range91,Doll90}.

In this paper we report the growth of single crystals as well as
the structural characterization of the \ZnMnInSe\ series. We used
X-ray techniques to study the transformation from the tetragonal
structure of \ZnInSe\ to the rhombohedral structure of \MnInSe.

\section{Experimental}
\subsection{Sample preparation}

Single crystals of \ZnMnInSe\ with nominal \Mn\ concentrations
$0{\le} x{\le}1$ were prepared by a vapour phase chemical
transport technique in an evacuated and sealed quartz tube of
$20$~$\mathrm{cm}$ length and $2$~$\mathrm{cm}$ diameter. The best
single crystals were obtained using $\mathrm{Al}\mathrm{Cl}_3$ for
high \Mn\ concentration compounds ($x>0.5$)\cite{Doll90} and
$\mathrm{I_2}$ for low \Mn\ concentration
($x<0.5$)\cite{Sagredo98, Schafer64} as transporting agents in the
reaction. About $5$~$\mathrm{mg/cm^3}$ of
$\mathrm{Al}\mathrm{Cl}_3$ ($4$~$\mathrm{mg/cm^3}$ $\mathrm{I_2}$)
was added into the ampoules together with $1.5$~$\mathrm{g}$ of
reactants. The starting materials for the growth were
polycrystalline samples prepared in a vertical furnace at
$1000^\mathrm{o}$$\mathrm{C}$.

The transport reaction was carried out in a two temperature zone
furnace in temperature gradients between $900$ and
$950^\mathrm{o}\mathrm{C}$ for $\mathrm{Al}\mathrm{Cl}_3$ and
between $800$ and $850^\mathrm{o}\mathrm{C}$ for $\mathrm{I_2}$.
The temperatures were ramped up at $100^\mathrm{o}\mathrm{C}$ per
day. The reaction periods were two or three days, after which the
temperature was lowered during a period of about five days. The
resulting crystals were layered, had black and bright faces and
were very flexible. Their dimensions were up to
$1$~$\mathrm{cm^2}$, with thicknesses between $20$ and
$30$~$\mu\mathrm{m}$.

The resulting \Mn\ concentrations, $x$, were obtained from the
Curie constant, extracted from high temperature magnetic
susceptibility measurements. Their precision was around 5\%, but
the method assumes stoichiometric amounts of the other elements.
Additional composition analysis of the crystals was carried out
with a Shimadzu EDX-900 energy dispersive X-Ray fluorescence
spectrometer. The assumed stoichiometry and the \Mn\
concentrations obtained from magnetic measurements were confirmed
within 10\%.

\subsection{X-Ray measurements}

X-Ray Powder Diffraction (XRPD) patterns of the powdered samples
were recorded using a Rigaku powder diffractometer utilizing
\Ni-filtered \Cu-$\mathrm{K}_\alpha$ radiation
($20$~$\mathrm{mA}$, $40$~$\mathrm{kV}$) $\lambda {=}
1.5418$~$\mathrm{\AA}$ in step scanning mode
($0.05^\mathrm{o}/10\;\mathrm{s}$). Data collection was done for
$2\theta$ between $10$ and $80$ degrees.

Single crystal Laue diagrams were registered using \Cu\ radiation
($20$~$\mathrm{mA}$, $40$~$\mathrm{kV}$), in transmission mode,
recorded on an image plate ($100$~$\mathrm{mm}$ $\times$
$86$~$\mathrm{mm}$) with imaging distance of $30$~$\mathrm{mm}$
from the crystal. The exposition time was $30$~minutes.

\section{Results and discussion}
\subsection{XRPD profiles}
\Fref{fig:xrpd} shows the experimental XRPD profiles for all the
samples studied. Peaks for the pure \MnInSe\ $(x{=}1)$ and
$x{=}0.87$ samples, as well as for the \ZnInSe\ with $x{=}0.01$
and $x{=}0.25$ samples could be indexed assuming the expected
crystal structures described in the introduction. For the
intermediate concentration samples, peaks from both structures
could be discerned, from which we infer a two-phase mixture. In
what follows, we will refer to the \MnInSe\ structure as the
``rhombohedral'' phase, and the \ZnInSe\ structure as the
``tetragonal'' phase.

\begin{figure}
  \includegraphics[width=\textwidth]{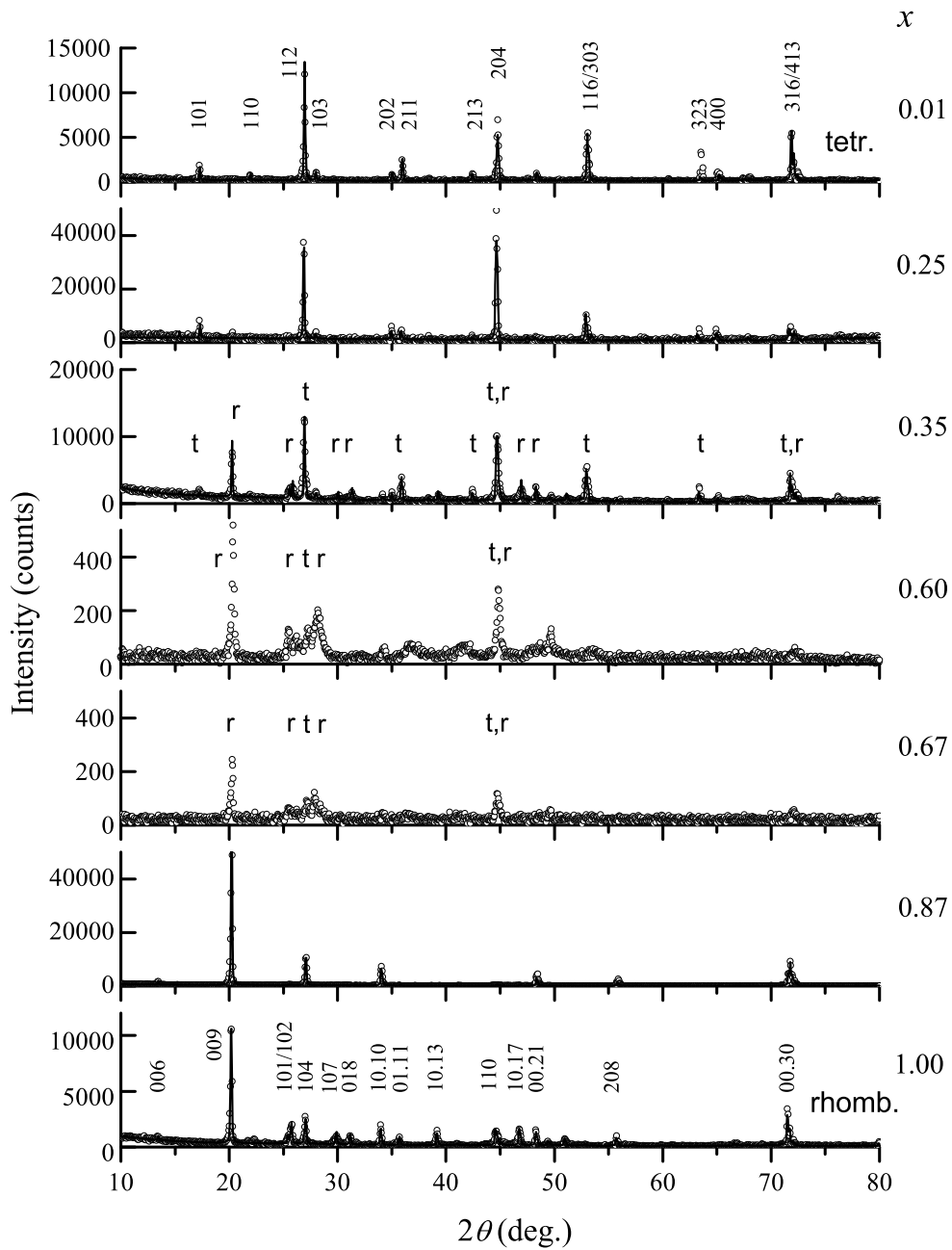}
\caption{\label{fig:xrpd}Experimental XRPD profiles (circles) for
the \ZnMnInSe\ samples. The continuous lines show the results of
Rietveld simulations. Also given are Miller indices for the peaks
in the tetragonal (\texttt{t}) $(x=0.01)$ and rhombohedral
(\texttt{r}) $(x=1)$ phases. For the two-phase samples the
identified peaks from each structure are indicated.}
\end{figure}


Starting from $x{=}1$ (\MnInSe), a substitution of 13\% of the
\Mn\ ions by \Zn\ ions $(x{=}0.87)$ results in the disappearance
of reflections from some crystallographic directions, while only
the strongest reflections are observed. Assuming the rhombohedral
structure is not completely destroyed, the surviving reflections
are mostly due to the hexagonal planes, perpendicular to the
$c$-direction of the unit cell. Since the Laue-diagram (see below)
still indicates a rhombohedral symmetry, we believe that at this
concentration we can still speak of a solid solution, albeit with
a decrease of the long range order in the structure.

The substitution with $x{=}0.67$ leads to an even more drastic
reduction of the number and intensities of the peaks, and to a
broadening of the profiles. This observation indicates that, at
this concentration, the presence of zinc affects the
crystallization of the compound even more. This is the least
crystalline of the samples we have investigated. Zinc substituting
\Mn\ in the rhombohedral phase causes strains in the crystal
network, which can lead to broaden reflections. Furthermore, one
observes that signs of the presence of a tetragonal phase begin to
appear. In \Fref{fig:xrpd}, the peaks corresponding to the
tetragonal and rhombohedral phases are labelled,  respectively,
``\texttt{t}'' and ``\texttt{r}'' above each observed reflection.
We suggest that in this sample the \Zn\ concentration has reached
its limit of solubility in the rhombohedral phase, and that the
tetragonal phase begins to segregate.

However, the $x{=}0.60$ sample becomes more crystalline, as one
can verify by the narrower peaks and their higher intensities.
More reflections due to the tetragonal phase begin to emerge, and
we now have clearly a two-phase system. This process is continued
with the $x{=}0.35$ sample, where one can notice the presence of
narrow and intense reflections corresponding to both the
rhombohedral and the tetragonal phases. This sample is more
crystalline than that with $x{=}0.60$, but now, one can infer the
segregation of the rhombohedral phase in a predominantly
tetragonal phase, since the sample is richer in \Zn\ atoms. When
the \Zn\ concentration increases to $75\%$ ($x{=}0.25$), only the
tetragonal phase is observed in the diffraction profiles. One can
suppose that the \Mn\ atoms are in solid solution within the
tetragonal phase. Finally, the crystal rich in zinc $(x{=}0.01)$
presents all the peaks expected for the \ZnInSe\ compound
\cite{Fiorani83}.

\subsection{Lattice parameters}
The XRPD patterns allowed the determination of the lattice
parameters. They were determined using single reflection peaks,
when possible, and pairs of peaks identified for each
sample\cite{ITC72}. The averages and error bars were calculated.
The results are shown in \Tref{table1}, together with literature
data for comparison.

\begin{table}
\caption{\label{table1} The lattice parameters obtained for each
observed phase in \ZnMnInSe. For the rhombohedral phase, $a$ and
$c$ refer to a hexagonal unit cell. For comparison, literature
standards are also listed.}
\begin{indented}
\item[]\begin{tabular}{@{}ccccc} \br
&\multicolumn{4}{c}{crystalline phase}\\
\cline{2-5}
&\multicolumn{2}{c}{tetragonal}&\multicolumn{2}{c}{rhombohedral}\\
\cline{2-5}
x (\%)& $a$ (\AA)& $c$ (\AA)& $a$ (\AA)& $c$ (\AA)\\
\mr
$0$ (standard)${}^{*}$&$5.710 {\pm} 0.001$&$11.420 {\pm} 0.002$&&\\
$1$&$5.719  {\pm} 0.014$&$11.514 {\pm} 0.048$&&\\
$25$&$5.736 {\pm} 0.009$&$11.517 {\pm} 0.042$&&\\
$35$&$5.729 {\pm} 0.006$&$11.489 {\pm} 0.023$&$4.044 {\pm} 0.009$&$39.444 {\pm} 0.008$\\
$60$&$5.694 {\pm} 0.015$&$11.418 {\pm} 0.096$&$4.064 {\pm} 0.015$&$39.275 {\pm} 0.078$\\
$67$&&&&$39.417 {\pm} 0.037$\\
$87$&&&$4.046 {\pm} 0.025$&$39.521 {\pm} 0.026$\\
$100$&&&$4.046 {\pm} 0.022$&$39.555 {\pm} 0.034$\\
$100$ (standard)${}^{**}$&&&$4.051 {\pm} 0.001$&$39.464 {\pm} 0.002$\\
\br
\end{tabular}\\
${}^{*}$(\ZnInSe, ICSD collection code 256470)\\
${}^{**}$(\MnInSe, ICSD collection code 69696)
\end{indented}
\end{table}

\begin{figure}
\begin{center}
\includegraphics[scale=0.45]{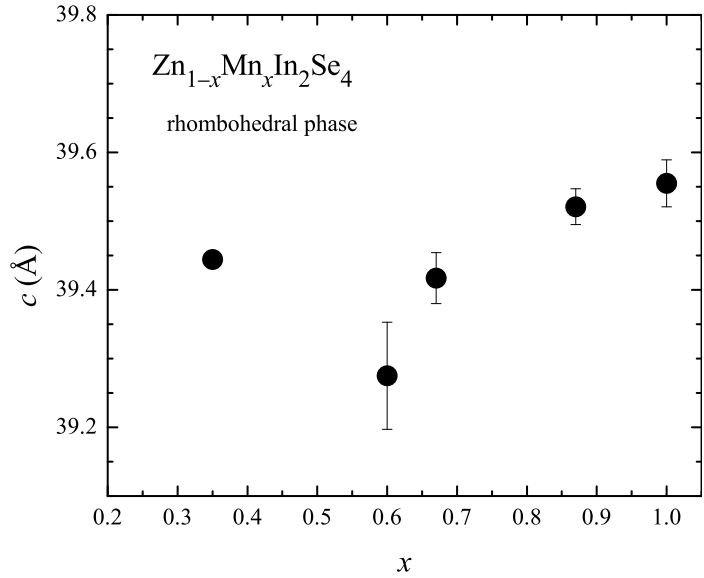}
\end{center}
\caption{\label{fig:param}Cell parameter $c$ of the rhombohedral
phase for samples with $x\geq0.35$. The decrease of $c$ for $x$
from 1 down  to $0.60$ is due to the smaller ionic radius of
$\mathrm{Zn^{2+}}$ compared to $\mathrm{Mn^{2+}}$. }
\end{figure}

For the samples with an average \Mn\ concentration $x\ge0.35$,
which contain the rhombohedral phase, the parameter $a$ does not
change considering the error bars. On the other hand, the
parameter $c$ decreases with the decrease of \Mn\ concentration
down to $x{=}0.60$ (See \Fref{fig:param}). The decrease of the $c$
parameter can be associated with the substitution of \Mn\ atoms in
the rhombohedral structure by \Zn\ atoms, since the ionic radius
of zinc is smaller than that of manganese. For the even higher
dilution $x{=}0.35$, the $c$ parameter increases. We interpret
this fact as due to the segregation of the more stable tetragonal
phase, leaving a smaller amount of \Zn\ atoms to go into the
rhombohedral phase. The cell parameters obtained for the
tetragonal phase for $x$ ranging from $0.35$ to $0.01$ are
practically constant considering the uncertainties.

The XRPD profiles were compared with Rietveld simulations using
the program PowderCell v2.4 \cite{Kraus98}. In the simulations the
previously obtained lattice parameters were introduced as
constants. The simulation procedure was carried out considering
the crystallites in the shape of plates and preferential
directions $[112]$ and $[001]$, respectively, for the tetragonal
and rhombohedral phases. The results are shown in \Fref{fig:xrpd}
as continuous lines over the experimental data points. The results
are in good agreement with the experimental data, indicating the
consistency of the interpretation.

\subsection{Laue Patterns}

\begin{figure}
\includegraphics{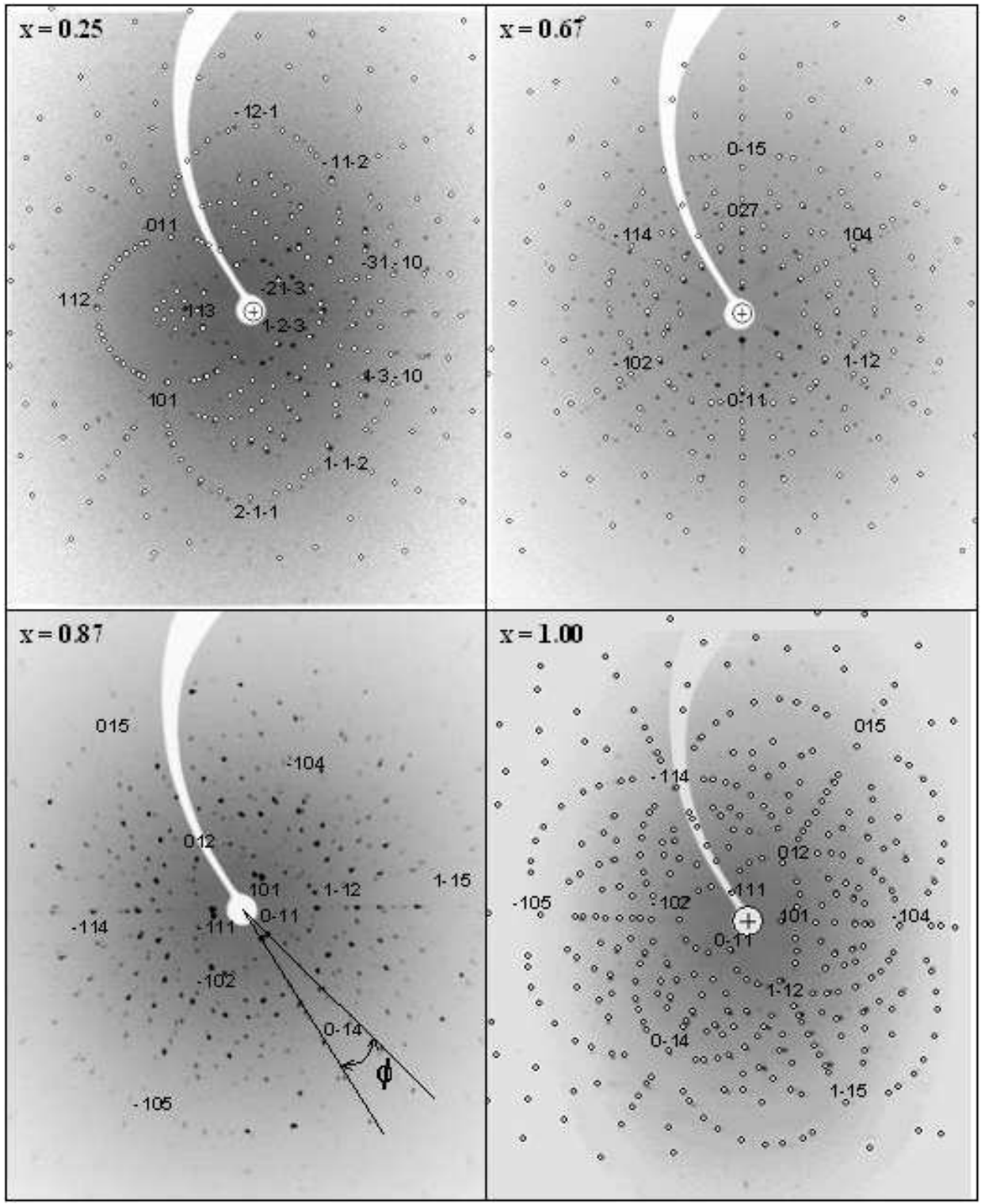}
\caption{\label{fig:laue}Laue diagrams for the samples with $x=1,
0.87, 0.67$ and $0.25$.
  White circles on the pictures for $x=0.25, 0.67$ and $1.00$ represent
  the OrientExpress simulations.\\
  }
\end{figure}
To further evaluate the phase transformation and the crystal
quality we measured Laue diagrams for the samples. It was possible
to collect the Laue patterns of the samples with $x {=} 1, 0.87,
0.67$ and $0.25$, which sizes were bigger than the cross-section
of the X-ray beam. The images were treated using OrientExpress
3.3\cite{LMGP}. The lattice parameters given in \Tref{table1} were
introduced in the data input files for the simulations. The method
used, based on the indexing of a small set of selected
reflections, proposes one or a small number of solutions. The
program computes and displays the corresponding simulated Laue
patterns (all reflections) or set of patterns. The best solution
is easily and unambiguously obtained through the visual comparison
of the experimental pattern with the set of simulated ones. Once
the correct solution is found, the program makes it possible to
compute the rotations which, applied to the sample holder axes,
will set the crystal to any desired new orientation. The
experimental images are displayed in \Fref{fig:laue}. For
comparison, the simulated Laue patterns and the indexes of some
reflecting planes are superposed over each image.

The $x{=}1$ sample (\MnInSe) presents the rhombohedral structure
and the data indicate that the crystal was grown towards the
$c$-axis, the preferential direction observed by XRPD. The good
agreement between the experimental and the simulated Laue patterns
points to a good crystal quality.

Consistent with our interpretation of the XRPD pattern, the sample
with $x{=}0.87$ also exhibits the rhombohedral structure. However,
this sample presents double reflections, which are rotated by
$\phi{\simeq}12^{\mathrm{o}}$ around the beam direction. This fact
can be associated with rotated planes (around the $c$-axis),
probably caused by distortions induced by zinc atoms substituting
the manganese atoms in the structure.

The Laue diagram of the sample with $x{=}0.67$ again shows the
symmetry expected for the rhombohedral phase and confirms the
interpretation of the XRPD data. However, the observed Laue
pattern cannot be reproduced by using a single orientation of the
crystal. For the displayed simulated pattern, the $c$-axis was
considered as the direction of the X-ray beam. The experimental
pattern can be simulated as the superposition of several crystals,
each with a different orientation for the $c$-axis. The relative
tilting of different crystals was up to about $9^{\mathrm{o}}$.
This observation points to a distortion of the rhombohedral
crystal structure along this axis in agreement with the line
broadening observed by XRPD\cite{Warren50,Harrison65}. Simulations
considering the presence of the tetragonal phase (as detected in
the XRPD spectrum) were made, but no signs of such a phase were
seen in the experimental Laue image. Apparently the minority
tetragonal phase segregates in a rhombohedral matrix, in the form
of small crystallites. They are not oriented coherently enough to
form a Laue image.

For the sample with $x{=}0.25$ the Laue pattern shows only the
tetragonal structure and the simulation indicates that the crystal
grew along the $[112]$ crystallographic direction. Accordingly,
the XRPD presents a more intense peak for this direction. The
$(112)$ crystal plane was oriented perpendicular to the X-ray beam
to obtain the Laue diagram for this sample.

\section{Conclusions}
X-ray diffraction measurements were performed on CVT grown
crystals of \ZnMnInSe, with $x$ ranging from $0.01$ to $1$. The
results indicate that the crystals present a purely rhombohedral
phase for $x\geq0.87$ and a purely tetragonal phase for
$x\leq0.25$. For the samples with $x$ between $0.67$ and $0.35$, a
mixture of rhombohedral and tetragonal phases was observed. These
results represent approximate limits on the range of
concentrations for which single phase solid solutions can be
grown. Furthermore, substitution of even small amounts of \Mn\
($x{=}0.25$) or \Zn\ ($x{=}0.87$) leads to a distortion of the
original structures and degraded crystallinity.

\ack
The authors thank  M. C. Fantini for helpful discussions.
This work was supported in part by the Brazilian agencies CNPq and FAPESP.

\end{document}